\documentclass[11pt,a4paper]{article}

\usepackage{inputenc}
\usepackage{amsmath}
\usepackage{bm}
\usepackage{bbold}
\usepackage{amsthm}

\usepackage{curves}
\usepackage{epic}

\usepackage{hyperref}

\title{On Parallel Implementation of a Discrete Optimization Random Search Algorithm\thanks{WSEAS Transactions on Computers, 2005, Vol.~4, No~9, pp.~1122-1129.}} 

\author{Nikolai~K.~Krivulin\thanks{Faculty of Mathematics and Mechanics, St.~Petersburg State University, 28 Universitetsky Ave., St.~Petersburg, 198504, Russia, 
nkk@math.spbu.ru.} \thanks{The work was partially supported by the NATO under Expert Visit Grant SST.EV.980134.}
\and
Dennis Guster\thanks{G.R.~Herberger College of Business, St.~Cloud State University, 720 4th Ave. South, St.~Cloud, MN 56301, USA, dcguster@stcloudstate.edu}
\and
Charles Hall\thanks{G.R.~Herberger College of Business, St.~Cloud State University, 720 4th Ave. South, St.~Cloud, MN 56301, USA, chall@bcrl.stcloudstate.edu}
}

\date{}

\begin{document}

\maketitle

\begin{abstract}
A random search algorithm intended to solve discrete optimization problems is considered. We outline the main components of the algorithm, and then describe it in more detail. We show how the algorithm can be implemented on parallel computer systems. A performance analysis of both serial and parallel versions of the algorithm is given, and related results of solving test problems are discussed.\\

\end{abstract}

\section{Introduction}

Together with other global optimization techniques \cite{Horst1990Global,Horst1995Handbook,Potter1994Acooperative,Zhigljavsky1991Theory}, including Simulated Annealing, evolutionary algorithms, and the Tunneling method, random search presents a powerful approach to solving discrete optimization problems when the objective function is too complex to obtain the solution analytically, or does not have an appropriate analytical representation. One can consider the functions with their values being obtained as a response from a controllable real-time process, or being evaluated through computer simulation. The optimization problems become even more difficult to solve if the evaluation of the function presents a very time-consuming procedure as is normally the case when the function is determined via computer simulation.

To solve the difficult optimization problems above, we propose a random search technique based on the Branch and Probability Bound (BPB) approach introduced in \cite{Zhigljavsky1985Mathematical} and further developed in \cite{Zhigljavsky1989Optimization,Zhigljavsky1991Theory,Zhigljavsky1993Semiparametric}. The BPB approach actually combines the usual branch and bound search scheme with statistical procedures of parameter estimation based on random sampling.

The key feature of the BPB approach is that it allows one to examine several regions within a feasible set concurrently in a natural way. Therefore, when solving multiextremal optimization problems, BPB algorithms normally offer an advantage over other global optimization techniques, which concentrate the search only on a single feasible region, and so could easily miss the solution.

If the sampling procedure including evaluation of the objective function at the sample points takes much more time than the core part of a search algorithm, it is quite natural to arrange the procedure so that it could work in parallel.

In this paper, we present a BPB random search algorithm together with its parallel implementation. A performance analysis of the parallel implementation is given based on solution of some test problems. As our computational experience shows, the parallel algorithm has a quite good potential to speedup the solution time when evaluation of the objective function is time-consuming.

\section{Problem and Solution Approach}

We consider the problem of finding
$$
x_{*}=\arg\min_{x\in X}f(x),
$$
where $ X $ is a discrete feasible set, and $ f $ is a real-valued function. As examples of $ X $, one can consider the set of integer vectors $ x=(x_{1},\ldots,x_{n}) $ with their components $ x_{i}\in\{1,\ldots,m\} $ for each $ i=1,\ldots,n $, or the set of all permutations from the permutation group of order $ n $.

It is assumed that the function value $ f(x) $ is available for each point $ x $ from the feasible set $ X $. However, we deal with the problems when the function itself may not have an analytical representation as it is usually the case in the analysis of outcome of actual real-time processes or output of computer simulation runs. Note that in the last case, the evaluation of the function may be a time-consuming procedure.

To solve the problems, we propose a global random search algorithm based on the BPB approach. As in the standard branch and bound scheme, the BPB approach involves partitioning the feasible set into subsets followed by choosing the subsets most promising for the solution. However, it assumes both partitioning and determining the subsets for further search to be performed on the basis of some statistical procedures. 

As with many other adaptive random search techniques, the BPB algorithms actually employ random sampling with both the feasible set and the sample probability distribution over the set being modified with each new iteration designed to exploit information about the function behavior, obtained in the course of the previous search. 

The BPB approach offers a natural and an efficient technique to control the search, based on a statistical procedure which estimates the prospectiveness of each subset for further consideration. The procedure evaluates a criterion based on sample data, which has a two-fold implementation. It allows one to reduce the feasible set by removing the subsets that have a low criterion, and so could hardly contain the solution. 

On the other hand, evaluation of the criterion plays the key role in rebuilding the sample distribution. In fact, the new distribution is defined in such a way that it provides for more intensive sampling resulting in more promising subsets with higher values of the prospectiveness criterion.

\section{Basic Components of the Algorithm}

In this section, we outline the basic concepts and components of the BPB random search algorithm.

\subsection{Prospectiveness Criterion}

Evaluation of prospectiveness of a subset for further search provides the basis of BPB algorithms. Consider a prospectiveness criterion introduced in \cite{Zhigljavsky1985Mathematical} (see also \cite{Zhigljavsky1991Theory,Zhigljavsky1993Semiparametric}). 

Let $ Z\subset X $ be a subset of the feasible set $ X $, $ {\Xi}=\{x_{1},\ldots,x_{K}\} $ be a sample from a probability distribution $ P(dx) $ over $ X $, and $ y_{*} $ is the minimum value of the function $ f $ over $ {\Xi} $:
$$
y_{*}=\min_{x\in{\Xi}}f(x).
$$

Assuming that $ {\Xi}\cap Z\ne\emptyset $, one can evaluate $ y=f(x) $ for each $ x\in{\Xi}\cap Z $ to obtain a sample $ {\Upsilon}=\{y_{1},\ldots,y_{N}\} $, where $ N=|{\Xi}\cap Z| $ is the cardinality of $ {\Xi}\cap Z $, and define $ y_{(1)}\leq\cdots\leq y_{(N)} $ to be the ordered statistics associated with $ {\Upsilon} $.

The prospectiveness criterion for the subset $ Z $ is defined as
\begin{equation}\label{E-Crit}
\varphi_{{\Xi}}(Z)
=\left(1-\left(\frac{y_{(1)}-y_{*}}{y_{(k+1)}-y_{*}}\right)^{\alpha}\right)^{k},
\end{equation}
where $ k $ is a positive integer number, and $ \alpha $ is a positive real parameter.

As it has been shown in \cite{Zhigljavsky1985Mathematical}, the criterion has a natural statistical interpretation. If $ k\to\infty $ and $ k^{2}/N\to0 $ as $ N\to\infty $, then $ \varphi_{{\Xi}}(Z) $ converges to the probability that 
$$
\min_{x\in Z}f(x)\leq y_{*}.
$$

In practice, the value of $ k $ can be set according to the following conditions. If $ N\geq10 $, then one can take
$$
k=\left\{
    \begin{array}{cl}
      \lfloor N/10\rfloor, & \mbox{if $ N<100$}, \\
      10,                  & \mbox{if $ N\geq100$};
    \end{array}
  \right.
$$
otherwise, one has to expand the sample $ {\Xi} $ until $ N=|{\Xi}\cap Z|\geq10 $, and then to try to evaluate the criterion once again.

The parameter $ \alpha $ is actually determined by the behavior of the function $ f $ on the entire feasible set $ X $, and it is normally unknown. To estimate $ \alpha $, suppose that $ y_{(1)}\leq\cdots\leq y_{(N)} $ are ordered statistics corresponding to the entire sample $ {\Xi} $ over $ X $. It can be shown \cite{Zhigljavsky1985Mathematical}, that the estimate
\begin{equation}\label{E-Alpha}
\widehat{\alpha}=
\ln5\left/\ln\frac{y_{(k+1)}-y_{(1)}}{y_{(m+1)}-y_{(1)}}\right.
\end{equation}
converges to $ \alpha $, if $ k\to\infty $, $ k^{2}/N\to0 $, and $ m/k\to0.2 $ as $ N\to\infty $. To evaluate $ \widehat{\alpha} $, one normally takes $ k=10 $, $ m=2 $, and $ N\geq100 $. 

Note that the above asymptotic results have been initially obtained under the assumption that the feasible set $ X $ is a compact subset of an Euclidean space. However, as our computational experience shows (see, e.g. \cite{Zhigljavsky1989Optimization}), the related practical recommendations still work well when solving optimization problems with discrete feasible sets.

\subsection{Representation of the Feasible Set}

At each iteration of the algorithm, the current feasible set $ X $ is represented as $ X=Z_{1}\cup\cdots\cup Z_{k} $, where $ Z_{j} $, $ j=1,\ldots, k $, are subsets of a common simple structure. The basic subset type, hyperballs or hypercubes with respect to a metric $ \rho $ are normally taken to provide for efficient partitioning and sampling procedures. Since for some discrete spaces (e.g., permutation groups), the concept of a hypercube is not appropriate, we restrict ourselves to hyperballs
$$
B_{r}(z,\rho)=\{x|\rho(z,x)\leq r\},
$$
where $ r $ is the radius, and $ z $ is a center.

Starting with a hyperball $ Z=B_{r}(z,\rho) $ of a radius $ r=R $, where $ R $ is large enough to cover the initial set $ X $ at the first iteration, the algorithm consecutively decrements the radius of hyperballs with every new iteration so as to allow for reduction of the feasible set and thereby concentrating the search on more promising subsets.

\subsection{Reduction and Partition of the Sets}

The reduction procedure is based on the partitioning of the current feasible set $ X $ into a subset $ Z $ and its complement $ X\setminus Z $.

In order to decide, if the complement can be removed, the procedure first evaluates its related criterion (\ref{E-Crit}) to get $ \gamma=\varphi_{\Theta}(X\setminus Z) $, where $ {\Theta} $ is the set of all sample points currently available. If the value of $ \gamma $ appears to be less than a fixed low bound $ \delta $, which determines the lowest level for subsets to be considered as candidates for further search, then the complement is removed.

The procedure actually combines reduction of the feasible set and partition of the reduced set into subsets, and can be described in more detail as follows. Suppose that $ r $ is the common radius of hyperballs, and $ \delta $ is the low bound for the criterion (\ref{E-Crit}). Let us define
$$
z_{1}=\arg\min_{x\in{\Theta}}f(x),
\qquad
Z_{1}=B_{r}(z_{1},\rho),
$$
and consider the value of $ \gamma_{1}=\varphi_{\Theta}(X\setminus Z_{1}) $.

If $ \gamma_{1}<\delta $, then the subset $ X\setminus Z_{1} $ can be removed since it has a very low prospectiveness level. Otherwise, when $ \gamma_{1}\geq\delta $, the procedure has to be continued. Now we take
$$
z_{2}=\arg\min_{x\in{\Theta\setminus Z_{1}}}f(x),
\qquad
Z_{2}=B_{r}(z_{2},\rho).
$$

After evaluation of $ \gamma_{2}=\varphi_{\Theta}(X\setminus(Z_{1}\cup Z_{2})) $ the procedure may be continued or ended depending on the value of $ \gamma_{2} $. If continued, the procedure is repeated as long as there is a subset to remove.

It may appear that there are not enough sample points available to evaluate the criterion. In this case, one has to stop the procedure, go back to extend the sample $ {\Theta} $, and then start the procedure from the beginning.

Suppose that the procedure is repeated $ k $ times before meeting the condition of removing a subset.
Upon completion of the procedure, we have the current feasible set $ X $ reduced to the union $ Z_{1}\cup\cdots\cup Z_{k} $, and the current set of sample points $ {\Theta} $ reduced to $ {\Theta}\cap(Z_{1}\cup\cdots\cup Z_{k}) $.

\subsection{Sample Probability Distribution}

To make a decision on how to reduce the current feasible set, the algorithm implements a statistical criterion based on random sampling over the set. This makes the sampling procedure a key component of the algorithm. The procedure applies a probability distribution, which is first set to the uniform distribution over the initial feasible set $ X $, and then modified with each new iteration.

Suppose that the current set $ X $ is formed by $ k $ subsets (hyperballs): $ X=Z_{1}\cup\cdots\cup Z_{k} $. The distribution $ P(dx) $ over $ X $ can be defined as a superposition of a probability distribution over the set of hyperballs and the uniform distribution over each hyperball. With a probability $ p_{j} $ assigned to the hyperball $ Z_{j} $, we have
$$
P(dx)=\sum_{j=1}^{k}p_{j}Q_{j}(dx),
$$
where $ Q_{j}(dx) $ denotes the uniform distribution over $ Z_{j} $ for each $ j=1,\ldots,k $.

The algorithm sets probabilities $ p_{1},\ldots,p_{k} $ in proportion to the criterion (\ref{E-Crit}) determined by their related hyperballs. In this case, the probabilities actually control the search, allowing the algorithm to put more new sample points into the hyperballs with higher probabilities.

In order to get the probabilities, one can evaluate $ q_{j}=\varphi_{\Theta}(Z_{j}) $ for each $ j=1,\ldots, k $, and then take
$$
p_{j}=q_{j}\bigg/\sum_{m=1}^{k}q_{m}.
$$

Note that there may be not enough sample points available when evaluating $ q_{j} $. In this case, it is quite natural to set $ q_{j}=\delta $. If it appears that all $ q_{j} $ equal $ 0 $, one can set $ q_{j}=1 $ for every $ j=1,\ldots,k $.

\section{Random Search BPB Algorithm}

Now we summarize the ideas described above in the presentation of the entire search algorithm.

The algorithm actually offers both global and local search capabilities. With each new iteration of the global search, the algorithm decrements the radius of the hyperballs by $ 1 $ until the radius achieves $ 1 $. All further iterations are performed with the radius fixed at $ 1 $ until a local minimum is found. We consider the best sample point found as a local minimum if all its nearest neighbors have already been examined, and are so included in the current set of sample points. 

\subsection*{Algorithm 1.}

\newcounter{Step}
\begin{list}{Step~\arabic{Step}.}{\usecounter{Step}}

\item Fix values for $ K $, $ R $ and $ \delta $. Set $ i=1 $, $ r_{0}=R $, 
$ {\Gamma}_{0}=\emptyset $, $ X_{1}=X $, and $ P_{1}(dx) $ to be the uniform distribution over $ X_{1} $. 

\item\label{S-2} Get a sample $ {\Xi}_{i}=\{x_{1}^{(i)},\ldots,x_{K}^{(i)}\} $ from $ P_{i}(dx) $. For each $ x\in{\Xi}_{i} $, evaluate $ f(x) $.

\item Set $ {\Theta}_{i}={\Gamma}_{i-1}\cup{\Xi}_{i} $, and find 
$$
y_{*}^{(i)}=\min_{x\in{\Theta}_{i}}f(x), \qquad
x_{*}^{(i)}=\arg\min_{x\in{\Theta}_{i}}f(x).
$$

\item If $ i=1 $, then evaluate $ \widehat{\alpha} $ with (\ref{E-Alpha}).

\item Put $ r_{i}=\max\{r_{i-1}-1,1\} $. 

\item If $ r_{i}=1 $ and $ B_{1}(x_{*}^{(i)},\rho)\subset {\Theta}_{i} $, then STOP.

\item Set $ k=1 $, $ U_{0}^{(i)}=\emptyset $.

\item\label{S-9} Find 
$ z_{k}^{(i)}=\displaystyle{\arg\min_{x\in{\Theta}_{i}\setminus U_{k-1}^{(i)}}f(x)} $.

\item Set $ Z_{k}^{(i)}=B_{r_{i}}(z_{k}^{(i)},\rho) $, $ U_{k}^{(i)}=U_{k-1}^{(i)}\cup Z_{k}^{(i)} $.

\item If $ |{\Theta}_{i}\cap (X_{i}\setminus U_{k}^{(i)})|\geq10 $, then evaluate 
$ \gamma_{k}^{(i)}=\varphi_{{\Theta}_{i}}(X_{i}\setminus U_{k}^{(i)}) $. Otherwise, replace $ {\Gamma}_{i-1} $ with $ {\Theta}_{i} $, and go to Step~\ref{S-2}.

\item If $ \gamma_{k}^{(i)}\geq\delta $, then replace $ k $ with $ k+1 $, and go to Step~\ref{S-9}. 

\item Set $ X_{i+1}=U_{k}^{(i)} $.

\item Set $ {\Gamma}_{i}={\Theta}_{i}\cap U_{k}^{(i)} $.

\item For each $ j=1,\ldots,k $, evaluate 
$$
q_{j}^{(i)}
=
\left\{
 \begin{array}{cl}
  \varphi_{{\Gamma}_{i}}(Z_{j}^{(i)}), & \quad\mbox{if $ |{\Gamma}_{i}\cap Z_{j}^{(i)}|\geq10 $}, \\
  \delta, & \quad\mbox{otherwise}.
 \end{array}
\right.
$$

\item For each $ j=1,\ldots,k $, evaluate
$$
p_{j}^{(i)}=q_{j}^{(i)}\bigg/\sum_{m=1}^{k}q_{m}^{(i)}.
$$

\item Set
$ P_{i+1}(dx)=\displaystyle{\sum_{j=1}^{k}p_{j}^{(i)}Q_{j}^{(i)}(dx)} $.

\item Replace $ i $ with $ i+1 $, and go to Step~\ref{S-2}.

\end{list}

\section{Parallel Version of the Algorithm}

In many practical situations, generating of sample points and/or evaluation of the objective function at the points present a time-consuming procedure. Specifically, sampling procedures can take a lot of time when the feasible set is large and has a complex structure. As another illustration, one can consider the evaluation of an objective function as a response from a lengthy simulation run.

On the other hand, the sampling procedure intended to produce many sample points in a unified way can normally be split into independent routines each turning out a part of the sample. Therefore, when the sampling procedure takes sufficiently more time than the other steps of the algorithm, one can achieve higher performance by rearranging the procedure to work in parallel.

In fact, when designed to work in parallel, Algorithm~1 retains its overall description with the only difference being that it now performs sampling (Step~2 and 3) as a parallel procedure.

\section{Test Problems}

As the feasible set for test problems, we consider the set of integer vectors $ x=(x_{1},\ldots,x_{n}) $, where $ x_{i}\in\{1,\ldots,m\} $ for each $ i=1,\ldots,n $. The number $ |X|=m^{n} $ of elements in $ X $ is finite. Note however, that, in practice, it can be very large.

\subsection{Metrics on the Feasible Set}
Selection of a suitable metric is very important in insuring that random search procedures are efficient. First, the metric should effectively separate points which considerably differ, and group points which are similar according to the nature of the problem under consideration. On the other hand, the metric has to provide for efficient algorithms for generating random points according to the uniform distribution over some standard (elementary) sets like hyperballs or hypercubes.

Consider the following three metrics on $ X $:
\begin{eqnarray*}
\rho_{1}(x,y)
&=&
\sum_{i=1}^{n}(1-\delta_{x_{i}y_{i}}), \\
\rho_{2}(x,y)
&=&
\max_{1\leq i\leq n}|x_{i}-y_{i}|, \\
\rho_{3}(x,y)
&=&
\sum_{i=1}^{n}|x_{i}-y_{i}|,
\end{eqnarray*}
where $ \delta_{ij}=1 $, if $ i=j $, and $ \delta_{ij}=0 $, otherwise.

It is easy to determine the maximum distance between two points for each metric:
\begin{eqnarray*}
\max_{x,y\in X}\rho_{1}(x,y)
&=&
n, \\
\max_{x,y\in X}\rho_{2}(x,y)
&=&
m-1, \\
\max_{x,y\in X}\rho_{3}(x,y)
&=&
n(m-1).
\end{eqnarray*}

Clearly, the metric $ \rho_{3} $ can be considered as providing for more granularity since it leads to a greater variety of values in the hyperball radius. Assuming $ n>m $, the metric $ \rho_{1} $ can be ranked second with respect to the same property.

Among two metrics $ \rho_{1} $ and $ \rho_{3} $ the first one can offer a more simple and therefore more efficient sampling procedure when using hyperballs. Taking that into account, we use the metric $ \rho_{1} $ normally referred to as the Hamming distance.

\subsection{Uniform Probability Distributions}

Now we discuss how the uniform distribution over a hyperball determined by the Hamming distance can be modeled. First note that any hyperball $ B_{r}(z,\rho) $ can be represented as
$$
B_{r}(z,\rho)
=
S_{0}(z,\rho)\cup S_{1}(z,\rho)\cup\cdots\cup S_{r}(z,\rho),
$$
where $ S_{i}(z,\rho)=\{x|\rho(z,x)=i\} $ is a hypersphere for each $ i=0,1,\ldots,r $.

The sampling over a hyperball can be arranged as a two-stage procedure: (i) a hypersphere in the hyperball is selected according to some distribution over the hyperspheres, and then (ii) the uniform distribution on the hypersphere is modeled. Clearly, the probability assigned to a particular hypersphere must be proportional to the total number of points on the hypersphere.

Since any hypersphere of radius $ i $ contains
$$
N_{i}=|S_{i}(z,\rho)|=(m-1)^{i}{n\choose i}
$$
points, the random selection of a hypersphere in a hyperball of radius $ r $ can be performed as follows.

\subsection*{Algorithm 2.}

\begin{list}{Step~\arabic{Step}.}{\usecounter{Step}}

\item Fix $ n $, $ m $, and $ r\leq n $.

\item For each $ i=0,1,\ldots,r $, evaluate
$$
N_{i}=(m-1)^{i}{n\choose i}.
$$

\item Set $ N=N_{0}+N_{1}+\cdots+N_{r} $.

\item For each $ i=0,1,\ldots,r $, evaluate 
$$
P_{i}=\frac{1}{N}\sum_{j=0}^{i}N_{j}.
$$

\item Get a random number $ u $ from the uniform distribution over $ [0,1] $.

\item As the radius of the hypersphere, take $ j=\min\{i| P_{i}\geq u\} $.
\end{list}

Upon selection of a hypersphere $ S_{r}(z,\rho) $, one can generate a point $ x=(x_{1},\ldots,x_{n}) $ on the hypersphere according to the uniform distribution.

\subsection*{Algorithm 3.}

\begin{list}{Step~\arabic{Step}.}{\usecounter{Step}}

\item Set $ i=1 $, $ x=z $, $ M=\{1,\ldots,m\} $, and $ N_{0}=\{1,\ldots,n\} $. 

\item\label{S-A3-2} Get a random integer $ j $ from the uniform distribution over $ N_{i-1} $.

\item Set $ N_{i}=N_{i-1}\setminus\{j\} $.

\item Get a random integer $ k $ from the uniform distribution over $ M\setminus\{z_{j}\} $.

\item Set $ x_{j}=k $.

\item Replace $ i $ with $ i+1 $. If $ i\leq r $ then go to Step~\ref{S-A3-2}.

\end{list}

\subsection{Test Functions}

To test both serial and parallel versions of the algorithm, simple unimodal and multimodal functions with the known global minimum are considered (see, e.g., \cite{Potter1994Acooperative} for more examples). We assume the functions to be defined on the set $ X=\{(x_{1},\ldots,x_{n})|x_{i}\in\{1,\ldots,m\}, 1\leq i\leq n\} $, provided that $ m $ is even, and $ m<n $.

First, we consider an integer analog of the De Jong's function:
\begin{equation}\label{D-FUN1}
f(x)=\sum_{i=1}^{n}(x_{i}-m/2)^{2}.
\end{equation}

As it is easy to see, the function is unimodal with the minimum $ f(x_{*})=0 $ achieved at the point $ x_{*}=(m/2,\ldots,m/2) $. 

The following integer function is of the Rastrigin type:
\begin{equation}\label{D-FUN2}
f(x)=nm+\sum_{i=1}^{n}\Big[(x_{i}-m/2)^{2}
-m\cos(k\pi(x_{i}-m/2)/m)\Big],
\end{equation}
where $ k $ is an integer parameter.

If $ k=0 $, the function coincides with De Jong's function, and it is unimodal. As $ k $ increases, the function becomes multimodal.

It has the global minimum $ f(x_{*})=0 $, where $ x_{*}=(m/2,\ldots,m/2) $.

The function 
\begin{equation}\label{D-FUN3}
f(x)=\sum_{i=1}^{n}|x_{i}-m/2|+\sum_{i=1}^{n-1}|x_{i}-x_{i+1}|
+|x_{n}-x_{1}|
\end{equation}
has a local minimum $ f(x_{*}^{(i)})=n|i-m/2| $ at each point $ x_{*}^{(i)}=(i,\ldots,i) $, where $ i=1,\ldots,m $; and the global minimum $ f(x_{*})=0 $ that is achieved at $ x_{*}=(m/2,\ldots,m/2) $.

Clearly, the optimization problems with these functions can immediately be solved analytically, and, in fact, do not call for any sophisticated computational procedures. However, they could provide the basis for a preliminary performance analysis of the algorithm and for a prediction of its behavior when solving actual problems.

\section{Computational Experience}

Now we turn to the discussion of practical implementation of both serial and parallel algorithms, including test results and performance analysis.

\subsection{Software and Hardware Support}

To investigate the performance, both serial and parallel versions of the algorithm were coded in C++ under the Linux RedHat 8.0 operating system. The parallel code is based on LAM~6.5.9. implementation \cite{Lam2003Mpi} of the Message Passing Interface (MPI) communication standard \cite{Gropp1994Using}. 

The parallel application consists of two modules; first one intended to run on the master computer, and the second designed to support slave computers. The code running on the master controls the communication with the slaves, and performs all the steps of the algorithm except for the sampling procedure.

The master computer starts operating by establishing connections and broadcasting some general information, including the parameters $ n $ and $ m $ of the feasible set, among the slave computers. At each iteration of the algorithm, it sends requests to all slaves to produce samples. The request to a particular slave includes the current radius of hyperballs, and its own list of hyperball centers accompanied by the numbers of points to be generated in each hyperball. 

The sample points and their related values of the function are sent back to the master. Upon completion of the current iteration, the next iteration is initiated until the stop condition is met. 

The software was tested on a cluster of Intel Pentium II/ 500MHz/ 128Mb RAM/ 10Gb HDD computers with 100BaseTX 100Mbit LAN. 

\subsection{Serial Algorithm Analysis and Tests}

We begin with the results of testing a serial version of the algorithm code, which actually does not include any MPI support. A series of test runs were performed with the test functions defined on the feasible set $ X $ with $ n=200 $, $ m=50 $. The low bound $ \delta $ was set to $ 0.1 $.

By performing the tests, we first try to understand how both $ K $ and $ R $ can affect the time the algorithm takes to find the solution. 

Note that for a large sample sizes $ K $, the time to produce and utilize one sample becomes quite significant, and generally leads to increased total time. On the other hand, one can expect that a large $ K $ provides for more accurate statistical procedures that could reduce the overall number of samples, and hence the total solution time.

Clearly, for smaller initial radius $ R $, the number of algorithm steps with the current radius $ r>1 $ should decrease. In fact, a small $ R $ allows the algorithm to be more concentrated around the best points found in early steps. As this takes place, one can expect to reduce the total time when solving problems with simple unimodal objective functions. However, for more complicated multimodal functions, this time could become even larger because of a possible rise in the number of examined points, especially at $ r=1 $.

The results of evaluating the total solution time for the test functions with $ K $ being varied from 50 to 300 and $ R $ from 10 to 190 show that on average the algorithm takes less time when both $ K $ and $ R $ are within the range from 50 to 100.

Let $ S $ be the time spent generating the samples and evaluating the function (sampling time), and $ A $ be the time the algorithm takes to utilize the samples (algorithm time). The total solution time of the algorithm can be represented as
\begin{equation}\label{E-TS}
T_{S}=S+A.
\end{equation}

Finally, let us denote the total number of sample points examined by the algorithm during solution process, as $ N $, and define $ S_{1}=S/N $ and $ A_{1}=A/N $ to represent average sampling and algorithm time for one sample point. 

In Table~\ref{T-T1}, we present a brief summary of the test results for the serial algorithm for each test function. The summary actually includes the average times and numbers of examined points, calculated over the entire series of test runs. 
\begin{table}[!ht]
\begin{center}
\begin{tabular}{||c|c|c|c|c|c|c||}
\hline\hline
Test & & \multicolumn{3}{|c|}{Run time} & \multicolumn{2}{|c||}{Point time} \\
func- & $N$ & \multicolumn{3}{|c|}{(sec.)} & \multicolumn{2}{|c||}{(msec.)} \\
\cline{3-7} 
tion & & $T_{S}$ & $S$ & $A$ & $S_{1}$ & $A_{1}$ \\
\hline
(\ref{D-FUN1}) & 132994 & 212 & 147 & 66 & 1.10 & 0.50 \\
(\ref{D-FUN2}) & 129270 & 675 & 554 & 121 & 4.29 & 0.94 \\
(\ref{D-FUN3}) & 199244 & 914 & 343 & 570 & 1.72 & 2.86 \\
\hline\hline
\end{tabular}
\caption{Summary results for the test runs.}\label{T-T1}
\end{center}
\vspace{-1ex}
\end{table}

\subsection{Parallel Algorithm Analysis}

The total time the parallel algorithm takes to get the solution can be written as
\begin{equation}\label{E-TP}
T_{P}=S/p+A+C,
\end{equation}
where $ p $ is the number of slaves, $ C $ is the time the master spends on transmission of control/sample data to/from slaves (communication time).

Clearly, with (\ref{E-TS}) and (\ref{E-TP}) the speedup the parallel algorithm can achieve using one master and 
$ p\geq1 $ slave computers, can be represented as 
$$
\sigma(p)
=
\frac{T_{S}}{T_{P}}
=
\frac{S+A}{S/p+A+C}.
$$

Let us denote the average data transmission time for one sample point as $ C_{1} $. Assuming the amount of control data the master sends to be well below that of the sample data it receives, one can expect $ C_{1}\approx C/N $. Now we can write
\begin{equation}\label{E-SPEEDUP}
\sigma(p)
\approx
\frac{S_{1}+A_{1}}{S_{1}/p+A_{1}+C_{1}}.
\end{equation}

With (\ref{E-SPEEDUP}) one can examine the conditions required for the parallel algorithm to achieve a true speedup, and estimate actual speedup in particular problems. Specifically, in order to get a speedup $ \sigma>1 $, one should have 
$$
r
=
\frac{S_{1}}{C_{1}}
>
\frac{p}{p-1}.
$$

If the algorithm time $ A_{1} $ appears to be much less than both the sampling time $ S_{1} $ and the communication time $ C_{1} $, we have the speedup
$$
\widetilde{\sigma}(p)
=
\frac{S_{1}}{S_{1}/p+C_{1}}
=
\frac{rp}{r+p}.
$$

Since at a fixed $ r $ it holds that $ \widetilde{\sigma}(p)\rightarrow r $ as $ p\rightarrow\infty $, one can see that $ r $ presents the maximum asymptotic speedup of the parallel algorithm.

Note, however, that the actual speedup can be much lower than $ r $. It depends on the value of $ A_{1} $, and approaches 1 when $ A_{1} $ becomes sufficiently large. In addition, the simplified model (\ref{E-TP}) does not take into account many of the details of the actual network communication process, which could affect the speedup adversely, especially when the level of parallelism increases. 

\subsection{Parallel Algorithm Tests}
To evaluate expected speedup, we need an estimate of the average communication time $ C_{1} $.

Considering that with $ n=200 $, the data length for one point comprises 408 bytes, including $ 2n=400 $ bytes for the integer components of the related vector, and 8 bytes for the function value. Our computational experience shows that the average time to transmit the point data is approximately equal to 1 millisecond.

\begin{figure}[!ht]
\begin{center}
\setlength{\unitlength}{1mm}
\begin{picture}(82,55)

\newsavebox\curveone
\savebox{\curveone}(60,25)[b]
{\linethickness{1.5pt}\curve(
 3, 6.15,
 6, 7.81,
 9, 8.58,
 12, 9.02,
 15, 9.31,
 18, 9.51,
 21, 9.66,
 24, 9.78,
 27, 9.87,
 30, 9.95,
 33, 10.01,
 36, 10.06,
 39, 10.11,
 42, 10.15,
 45, 10.18,
 48, 10.21,
 51, 10.24,
 54, 10.26,
 57, 10.28,
 60, 10.30
)}

\newsavebox\curvetwo
\savebox{\curvetwo}(60,25)[b]
{\linethickness{1.5pt}\curvedashes{1,2}\curve(
 3, 8.39,
 6, 12.80,
 9, 15.52,
 12, 17.37,
 15, 18.70,
 18, 19.71,
 21, 20.50,
 24, 21.13,
 27, 21.65,
 30, 22.09,
 33, 22.46,
 36, 22.78,
 39, 23.06,
 42, 23.30,
 45, 23.51,
 48, 23.70,
 51, 23.87,
 54, 24.03,
 57, 24.17,
 60, 24.29
)}

\newsavebox\curvethree
\savebox{\curvethree}(60,25)[b]
{\linethickness{1.5pt}\curvedashes{1,1}\curve(
 3, 8.21,
 6, 9.71,
 9, 10.34,
 12, 10.68,
 15, 10.90,
 18, 11.05,
 21, 11.16,
 24, 11.24,
 27, 11.31,
 30, 11.36,
 33, 11.41,
 36, 11.45,
 39, 11.48,
 42, 11.51,
 45, 11.53,
 48, 11.55,
 51, 11.57,
 54, 11.58,
 57, 11.60,
 60, 11.61
)}

\put(15,12){\line(1,0){66}}
\put(15,12){\line(0,1){36}}

\put(81,48){\line(-1,0){66}}
\put(81,48){\line(0,-1){36}}

\multiput(18,12)(15,0){5}{\line(0,-1){2}}

\put(17,6){$0$}
\put(32,6){$5$}
\put(46,6){$10$}
\put(61,6){$15$}
\put(76,6){$20$}

\put(33,0){\bf Number of Slaves}

\multiput(15,15)(0,10){4}{\line(-1,0){2}}
\multiput(15,20)(0,10){3}{\line(-1,0){1}}

\put(7,14){$0.0$}
\put(7,24){$1.0$}
\put(7,34){$2.0$}
\put(7,44){$3.0$}

\put(0,52){\bf Speedup}

\put(73,42){$(\ref{D-FUN2})$}
\put(73,30){$(\ref{D-FUN3})$}
\put(73,20){$(\ref{D-FUN1})$}

\put(18,15){\makebox(60,25){\usebox{\curveone}}}
\put(18,15){\makebox(60,25){\usebox{\curvetwo}}}
\put(18,15){\makebox(60,25){\usebox{\curvethree}}}

\end{picture}
\caption{Predicted speedup for the test functions.}\label{F-F1}
\end{center}
\end{figure}
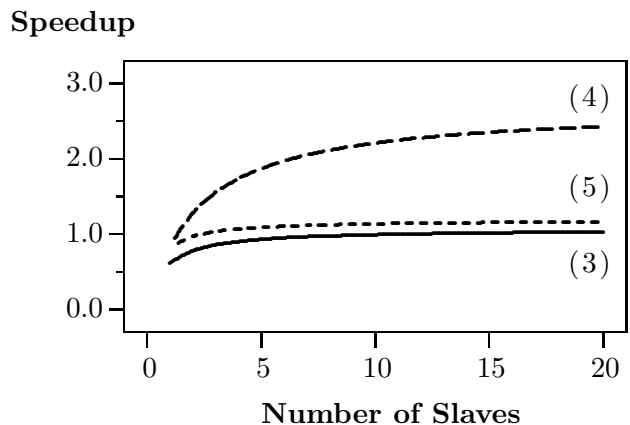

With $ C_{1}\approx1 $, and parameters $ S_{1} $ and $ A_{1} $ taken from Table~\ref{T-T1}, one can apply (\ref{E-SPEEDUP}) to evaluate the speedup for any $ p\geq1 $ (see Fig.~\ref{F-F1}). 

As it easy to see, one can expect an actual speedup only for the function (\ref{D-FUN2}) having the best value of 
$ r=S_{1}/C_{1}\approx 4.29 $. For the other functions, any sufficient speedup can hardly be achieved because of the low level of $ r=1.10 $ for (\ref{D-FUN1}), and a high magnitude of $ A_{1}=2.86 $ for (\ref{D-FUN3}).

\begin{figure}[!ht]
\begin{center}
\setlength{\unitlength}{1mm}
\begin{picture}(82,57)

\newsavebox\piecewiseone
\savebox{\piecewiseone}(60,25)[b]
{\linethickness{1.5pt}\dottedline{1.25}%
(0, 24.72)
(12, 30.03)
(24, 15.25)
(36,  9.50)
(48, 12.89)
(60,  8.69)
}

\newsavebox\piecewisetwo
\savebox{\piecewisetwo}(60,25)[b]
{\thinlines\drawline
(0, 19.44)
(12, 29.02)
(24, 10.75)
(36,  6.06)
(48,  7.06)
(60, 15.30)
}

\newsavebox\piecewisethree
\savebox{\piecewisethree}(60,25)[b]
{\linethickness{1.5pt}\dottedline{0.25}%
(0, 18.80)
(12, 31.58)
(24,  9.53)
(36,  4.96)
(48,  5.08)
(60,  7.21)
}

\put(15,12){\line(1,0){66}}
\put(15,12){\line(0,1){38}}

\put(81,50){\line(-1,0){66}}
\put(81,50){\line(0,-1){38}}

\multiput(18,12)(12,0){6}{\line(0,-1){2}}

\put(17,6){$0$}
\put(29,6){$1$}
\put(41,6){$2$}
\put(53,6){$3$}
\put(65,6){$4$}
\put(77,6){$5$}

\put(33,0){\bf Number of Slaves}

\multiput(15,15)(0,8){5}{\line(-1,0){2}}

\put(7,14){$400$}
\put(7,22){$500$}
\put(7,30){$600$}
\put(7,38){$700$}
\put(7,46){$800$}

\put(0,53){\bf Total Time (sec.)}

{\linethickness{1.5pt}\dottedline{1.25}(55,45)(65,45)}
\put(68,44){50}
{\thinlines\drawline(55,40)(65,40)}
\put(68,39){100}
{\linethickness{1.5pt}\dottedline{0.5}(55,35)(65,35)}
\put(68,34){150}

\put(18,15){\makebox(60,25){\usebox{\piecewiseone}}}
\put(18,15){\makebox(60,25){\usebox{\piecewisetwo}}}
\put(18,15){\makebox(60,25){\usebox{\piecewisethree}}}

\end{picture}
\caption{Average total solution time.}\label{F-F2}
\end{center}
\end{figure}
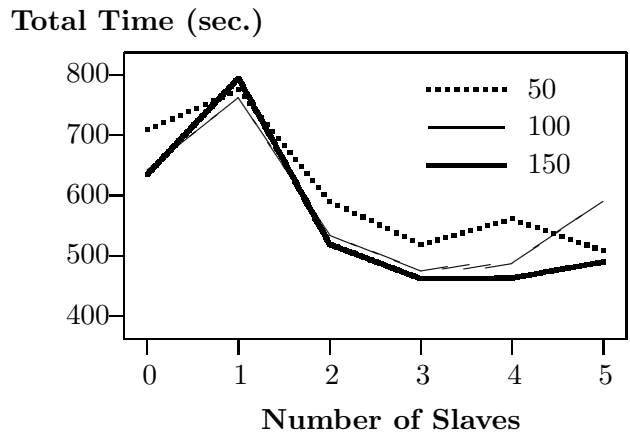

In order to evaluate actual speedup for function (\ref{D-FUN2}), several series of test runs were performed for each $ K=50,100,150 $, and $ p=1,2,3,4,5 $. One series involves a particular run for every value of $ R $ varied from 50 to 150 by 10. The average total solution time over the values of $ R $ for each series is represented in Fig.~\ref{F-F2}, where $ p=0 $ corresponds to the serial version of the algorithm.

One can see that the best speedup achieved was about 1.6-1.7 when using one master and 3 slave computers. Although the speedup appears to be relatively small, it does demonstrate the potential of parallelization. Since evaluation of the functions involves only a few operations, the sampling procedure does not take much time to produce samples. As the performance analysis shows, if this procedure is time-consuming, one can expect to achieve even greater efficiency.

\bibliographystyle{utphys}

\bibliography{On_parallel_implementation_of_a_discrete_optimization_random_search_algorithm}

\end{document}